\documentclass[pra, twocolumn]{revtex4}

\usepackage{amsmath,amssymb,amscd}  
\usepackage{psfrag}  
\usepackage[dvips]{epsfig}  
\usepackage{epsfig}  
\usepackage{dcolumn}
\usepackage{multirow}

\def\figurelib{.}

\usepackage{enumerate}
\usepackage{color}

\usepackage{theorem}
\newtheorem{algorithm}{Algorithm}

\theoremstyle{break}

\def\QED{~\rule[-1pt]{5pt}{5pt}\par\medskip}

\def\CalH{\mathcal{H}}

\def\su{\mathfrak{su}}

\def\la{\langle}
\def\ra{\rangle}

\def\Eq{Eq.~\eqref}

\DeclareMathOperator{\tr}{Tr}

\DeclareMathOperator{\diag}{diag}

\newcommand{\ma}[1]{\left[\begin{matrix} #1 \end{matrix}\right]}

\newcommand{\Cmn}[2]{\mathbb{C}^{#1 \times #2}}

\newcommand{\ppfrac}[2]{\frac{\partial #1}{\partial #2}}

\newcommand{\bra}[1]{|{#1}\rangle}
\newcommand{\ket}[1]{\langle{#1}|}

\def\ie{{\it i.e.}}

\usepackage{color}

\begin{document}   
\title{Optimal Control for Electron Shuttling} 

\author{Jun Zhang$^{1,2}$, Loren Greenman$^2$, Xiaotian Deng$^2$, Ian
  M. Hayes$^2$, and K. Birgitta Whaley$^2$} \affiliation{$^1$Joint
  Institute of UM-SJTU, Shanghai Jiao Tong University, and Key
  Laboratory of System Control and Information Processing, Ministry of
  Education, Shanghai, 200240,
  China\\
  $^2$Department of Chemistry, Berkeley Center for Quantum Information
  and Computation, University of California, Berkeley, California
  94720, USA}

\date{\today}

\begin{abstract}  
  In this paper we apply an optimal control technique to derive
  control fields that transfer an electron between ends of a chain of
  donors or quantum dots. We formulate the transfer as an optimal
  steering problem, and then derive the dynamics of the optimal
  control. A numerical algorithm is developed to effectively generate
  control pulses. We apply this technique to transfer an electron
  between sites of a triple quantum dot and an ionized chain of
  phosphorus dopants in silicon.  Using the optimal pulses for the
  spatial shuttling of phosphorus dopants, we then add hyperfine
  interactions to the Hamiltonian and show that a 500 G magnetic field
  will transfer the electron spatially as well as transferring the
  spin components of two of the four hyperfine states of the
  electron-nuclear spin pair.
\end{abstract}  

\maketitle 

\section{Introduction}
The benefits of implementing a quantum computer in
silicon~\cite{Morton_Nature_Review}, namely the ability to exploit the
techniques of the semiconductor industry and long electron and nuclear
spin coherence times, has been offset with challenges including the
coupling of qubits.  One mechanism for exchanging quantum information
between qubits is electron shuttling, in which spin or charge qubits
are physically moved between local sites~\cite{Skinner_Proposal}.  For
dopant spin qubits in silicon~\cite{Kane_Proposal}, electron shuttling
has been proposed using voltage gates and pulses designed analogously
to the Stimulated Raman Adiabatic Passage (STIRAP)
procedure~\cite{Bergmann_RMP_98}; this procedure is referred to as
Coherent Tunneling by Adiabatic Passage
(CTAP)~\cite{Greentree:04,Hollenberg_PRB_06,Rahman:10}.  Similar
mechanisms have been suggested~\cite{Rahman:10,Chen:11} for quantum
dots, which have also been proposed as
qubits~\cite{Loss_Divincenzo_Proposal,Jacak_APPA_00} in
silicon~\cite{Culcer_PRB_09,Hada_JJAP_04,Maune_Nature_481_344_2012} as
well as other materials such as
GaAs~\cite{Petta_Science_05,Taylor_NP_05}.  CTAP and other adiabatic
procedures avoid populating undesired sites at any point during the
transfer, thereby eliminating issues of decoherence associated with a
specific site.  However, if the source of decoherence is not
site-specific or is controllable by alternative means, it may be
useful to approach the state transfer problem for silicon qubits using
optimal control theory instead.  Optimal controls which minimize
transfer time or (as will be explored here) minimize pulse fluence
have been shown to minimize decoherence due to additive and
multiplicative white noise,
respectively~\cite{Brif_inprep_11,Moore_PRA_86_062309_2012}.  Such
noise sources can arise from thermal fluctuations of carriers.
Another noise source can arise from variations in the devices due to
the manufacturing, this introduces ${1}/{f}$ noise, and it will be
seen that our fluence-minimized pulses are dominated by high-frequency
components.  Additionally, for qubits such as charge qubits in which
site-specific decoherence is not the main problem, we can use optimal
controls to minimize the energy required for gate operations.
  
In this paper we will investigate the shuttling of electrons between
the ends of a qubit chain using optimal control theory.  Depending on
whether the chain represents dopant spin
qubits~\cite{Greentree:04,Rahman:10} or lateral quantum
dots~\cite{Chen:11}, the control fields will affect the tunnel
couplings between dopants or the on-site energy of a quantum dot,
respectively.  The task is to design some appropriate control fields
to transport the electron to the end of the qubit chain.  During this
process, quantum information can be passed through along the array so
as to realize the desired quantum information processing in the
solid-state quantum bits.

We have formulated this problem as an optimal steering problem in
control theory in a state space of all of the density matrices.  The
system dynamics is governed by the Liouville-von Neumann equation,
\begin{equation}
\label{eq:35}
i \dot \rho=[H, \rho].
\end{equation}
We use density matrices to formulate the steering problem in order to
make the extension to open systems clear.  The Hamiltonian contains
several control terms that can be altered externally to guide the
system towards a desired result, as well as a fixed drift term that is
determined by the physical nature of the system. The objective of
steering is to find control fields that transfer the system from an
initial state to a final state at a finite terminal time. This is also
known as the {\it constructive controllability} problem~\cite{Sas99}.
  
One way to solve the constructive controllability problem is to impose
a cost function, which may steer towards a minimum pulse energy, a
shortest transfer time, or a minimum error sum over all time steps.
We can then employ standard optimal control techniques such as the
Pontryagin maximum principle~\cite{Pontryagin:62} to derive optimality
conditions for the control fields. 
  
In the current paper we derive the underlying dynamics that govern the
time evolution of the optimal control pulses that minimize the pulse
fluence.  We choose this cost function partially for mathematical
reasons, namely that it provides a numerically well-behaved set of
equations for this and other numerical algorithms including the Krotov
method~\cite{Krotov_ARC_60_1427_1999,Sklarz_PRA_66_053619_2002,Palao_PRA_68_062308_2003,Reich_JCP_136_104103_2012},
but also because this choice of cost function minimizes heating in the
devices which would lead to decoherence as well as to effects of
multiplicative white noise.  We take an intuitive approach, which
yields the same results as a more formal approach using the
Lie-Poisson reduction theorem~\cite{Marsden:98}. For a given initial
condition, the resulting dynamics completely determine the time
evolution trajectories of the control pulse.  Hence, to solve for
control fields that achieve the desired state transfer, we just need
to find an appropriate initial condition. Finding such initial
conditions thus becomes an optimization problem on real
finite-dimensional space.
  
To solve the Liouville-von Neumann equation~\eqref{eq:35} numerically,
we divide the total time into a number of steps and then use piecewise
constant functions to approximate time-varying control fields. The
fidelity of the achieved state,
  \begin{equation}
  \label{eq:16}
  {F}=\tr \rho_T\rho(T),
\end{equation}
  with $\rho_T$ representing the desired state,
   thus depends on all of
  the piecewise constant control values, which themselves are
  dependent on the initial conditions of the dynamics as discussed above. 
  Using the chain rule, we can obtain the gradient of the fidelity
  with respect to the initial condition in an explicit form.  With
  this approach we can implement  gradient 
  algorithms to solve for the initial conditions that lead to the optimal control fields.
  
  To exemplify this approach, we investigate here the electron
  shuttling problem for three-donor systems. Our control algorithm
  derivation can be readily extended to systems with more donors.  We
  demonstrate the efficacy of our control algorithm by applying it to
  two physical systems taken from Refs.~\cite{Greentree:04}
  and~\cite{Chen:11}, namely electron shuttling across a chain of
  quantum dots and across a chain of phosphorus donors implanted in
  silicon.

\section{Mathematical background and formulation}
In this section we summarize the mathematical representation of the
electron shuttling problem and introduce some necessary mathematical
background for an optimal control treatment of this.

We consider here physical devices in which the spatial location of the
electron may be represented by three
qubits~\cite{Greentree:04,Chen:11,Rahman:10}.  The physical systems of
interest in this work describe the shuttling of a single electron
between either three quantum dots or three donor ions.  In both cases,
the electron is moving between distinct spatial locations or
``sites''.  Formally, the presence or absence of the electron on a
given site is then represented by the state of a qubit indexed by that
site.  For shuttling across a chain of quantum dots, the qubit state
coding for presence of an electron corresponds to the state of the
electron in a discrete energy level of the quantum dot.  For shuttling
across a chain of phosphorus donors implanted in silicon, the qubit
state coding for presence of an electron corresponds to a neutral
donor atom, {\it i.e.}, the electron is bound to the phosphorus
nucleus at that site.  Within these simplified physical
representations of electron shuttling over three sites, the system
Hamiltonian is defined on the Lie algebra $\su(3)$, \ie, all $3\times
3$ skew-Hermitian matrices.  The dynamics of the electron are
determined by the Liouville-von Neumann equation~\eqref{eq:35}, with
$\rho \in \Cmn{3}{3}$ as the density matrix of the three-site system.
Note that the density matrix is a Hermitian matrix with unit trace.
The Hamiltonian $H$ can be written in a general form as (setting
$\hbar=1$)
\begin{equation}
  \label{eq:11}
 i H=i H_0+\sum_{l=1}^8 u_lX_l=\sum_{l=1}^{8} a_lX_l
+\sum_{l=1}^8 u_lX_l,
\end{equation}
where $H_0$ is the drift term, $u_l$ are control fields, and the
matrices $X_i$ define a basis for $\su(3)$,
\begin{equation}
  \label{eq:2}
    \begin{aligned}
X_1&=\ma{0&i&0\\i&0&0\\0&0&0},  &X_2&=\ma{0&0&0\\0&0&i\\0&i&0},\\
X_3&=\ma{0&0&1\\0&0&0\\-1&0&0}, &X_4&=\ma{0&1&0\\ -1&0&0\\0&0&0}, \\
X_5&=\ma{0&0&0\\0&0&1\\0&-1&0}, &X_6&=\ma{0&0&i\\0&0&0\\i&0&0}, \\
X_7&=\ma{i&0&0\\0&-i&0\\0&0&0}, &X_8&=\frac{1}{\sqrt{3}}\ma{i&0&0\\0&i&0\\0&0&-2i}.
  \end{aligned}
\end{equation}
Note that this basis is just a rearrangement of the Gell-Mann
matrices~\cite{Georgi:99}.  The density matrix equation of motion is
then determined by the Liouville-von Neumann equation~\eqref{eq:35} with
the Hamiltonian given in (\ref{eq:11}).

The desired high fidelity implementation of electron shuttling amounts
to designing control functions $u_l$ that transfer the density matrix
$\rho$ from the initial state
\begin{equation}
  \label{eq:6}
  \rho_0=\ma{1&0&0\\0&0&0\\0&0&0}
\end{equation}
at time $t=0$ to the final state
\begin{equation}
  \label{eq:7}
  \rho_T=\ma{0&0&0\\0&0&0\\0&0&1}
\end{equation}
at time $t=T$.  

\section{Optimal control formulation and numerical algorithm}
\label{sec:oc}
To solve the state transfer problem presented in
Eqs.~\eqref{eq:35},~\eqref{eq:11},~\eqref{eq:6},~and~\eqref{eq:7}, we
formulate an optimal control problem by imposing a cost function, and
then use Pontryagin's maximum principle~\cite{Pontryagin:62} to derive
the optimality conditions. Based on these conditions, we can develop
an effective numerical algorithm to solve for the values of the
control fields.

\subsection{Optimal control formulation}
\label{sec:lg_cc}
In a typical control problem, we apply external control fields to a
system with the expectation that it will evolve towards a desired
state or objective.  The transfer of a system from an initial state to
a desired final state is
often referred to as the {\it controllability} problem. The criteria
to determine controllability for a general nonlinear system were
studied in Refs.~\cite{Jur72,Brockett:73b}, and the extensions to
quantum mechanical systems were reported in
Refs.~\cite{Tarn:81,Huang:83}.
  
From the controllability analysis for control systems on Lie groups,
it can be concluded that the system is controllable provided that the
drift term $H_0$ and control terms $X_l$ in \Eq{eq:11} can generate
the Lie algebra $\su(3)$~\cite{Jurdevic:97,Sas99}.  However, such an
analysis gives us only an existence result; it does not tell us how to
generate the necessary control fields.  What we are more interested is
the constructive controllability, \ie, finding the controls that
realize the state transfer.
  
One method of solving this problem is to impose a cost function to the
state transfer problem and then apply an optimal control method such
as the Pontryagin maximum principle~\cite{Pontryagin:62}. 
This yields a set of differential equations which must be satisfied by
the control fields.  In the following section, we illustrate the
construction of these equations for the electron shuttling problem.
     
We seek the control fields that not only realize the desired state
transfer but also minimize the cost function, where the latter is
defined as the time integral of a running cost that depends on the
control fields $u$ (from now on, we use $u$ to denote the finite set
of control fields $u_l$):
\begin{equation}
  \label{eq:13}
 \int_0^T L(u) dt.
\end{equation}
The integrand $L$ in Eq.~\eqref{eq:13}, referred to as the {\it
  running cost}, can be chosen quite generally to suit different
control objectives. 
For example, when $L=1$, minimization of the cost function will correspond to minimum
time control.  Here we choose $L$ as a quadratic function of $u$, which allows
minimization of pulse fluence:
\begin{equation}
  L(u)=\sum _l u_l^2 .
  \label{eq:lg_fluence}
\end{equation}
This choice of cost function minimizes heating in the devices, which
can cause decoherence if left unchecked and has also been shown to
minimize errors due to multiplicative white
noise~\cite{Brif_inprep_11,Moore_arXiv_11}.  With time variable
control fields $u_l(t)$, the cost function is thus a functional of
$u$.  The optimal control fields are defined as those fields that
minimize the cost functional, Eq.~(\ref{eq:13}). The task of finding
the optimal control fields is then expressed mathematically as the
task of minimizing the cost functional with respect to all possible
variations in all $u_l(t)$, i.e., the optimal $u$ yields
\begin{equation}
  \label{eq:13a}
  \min_{u(\cdot)} \int_0^T L(u) dt.
\end{equation}  

We note that the major motivation to add a cost function at this point
is to apply optimal control theory to solve the constructive
controllability problem presented in the previous section. For a
control Hamiltonian that depends linearly on an unbounded control
function, optimal control theory may not be applicable to minimum time
control. This is avoided when the running cost $L$ is chosen to be a
quadratic function of the control fields $u$, which provides another
motivation for the current choice of $L(u)$.

There are several possible approaches to solve the resulting optimal
control problem.  One common method for obtaining numerical solutions
to optimal control for quantum systems is the Lagrangian formalism in
which a Lagrange multiplier is defined to allow the system dynamics,
Eq.~(\ref{eq:35}), to be combined with the cost function to create a
new cost functional which is then optimized by solving the associated
Euler-Lagrange critical equations~\cite{Peirce:88,Zhu:98,Maday:03}.
We employ here the alternative Hamiltonian approach based on the
Pontryagin maximum principle.  While for many physical systems of
interest the two approaches arrive at equivalent formulations of the
equations to be solved for the optimal solutions, these are generally
in the form of two point boundary value problems.  Numerical solution
of such problems often require significant computational power and
considerable efforts have been made to develop effective algorithms
for their
solution~\cite{Krotov_ARC_60_1427_1999,ohtsuki1999monotonically,
  Sklarz_PRA_66_053619_2002,Palao_PRA_68_062308_2003,ohtsuki2004generalized,doria2011optimal,Reich_JCP_136_104103_2012}.
In the present case however, the Hamiltonian approach of the
Pontryagin maximum principle allows for a novel formulation of the
optimization as an initial value problem on a finite dimensional space
defined by a number of momentum functions~\cite{Marsden:98}.  This
allows the optimal solutions to be obtained with a relatively
straightforward numerical algorithm.

In the Pontryagin approach~\cite{Bry+Ho75} we define a co-state matrix
$\Psi$ that plays the role of a conjugate variable to $\rho$ in a
(classical) control Hamiltonian 
\begin{equation}
  \label{eq:14}
    \begin{aligned}
\CalH &=\la \Psi,[-iH, \rho]\ra+L(u)\\
&=-\left\la \Psi,\left[\sum_{l=1}^{8} a_lX_l+\sum_{l=1}^8
       u_lX_l, \rho\right]\right\ra+L(u),
  \end{aligned}
\end{equation}
 where $\la X, Y \ra$ denotes the matrix inner product of $X$ and $Y$:
 \begin{equation}
  \label{eq:46}
  \la X, Y \ra=\tr(XY^\dag).
  \end{equation}
  The equations of motion for $\Psi$ and $\rho$ are then obtained from
  the Hamilton equations for $\CalH$:
  \begin{equation}
\label{eq:46a}
\begin{aligned}
  \dot\Psi &=-\frac{\partial\CalH}{\partial \rho} = -[iH, \Psi] \\
  \dot\rho & =  \frac{\partial\CalH} {\partial \Psi} = -[iH,\rho].
\end{aligned}
\end{equation}
From this it is evident that the co-state matrix $\Psi$ plays the
formal role of a momentum variable.  We distinguish this from the
momentum functions defined as~\cite{Marsden:98}
\begin{equation}
  \label{eq:12}
\phi_l=\la \Psi,[X_l,\rho]\ra.
\end{equation}
Dimensional analysis shows that in this case the momentum function
corresponds formally to a kinetic energy function determined by $\rho$
and its conjugate variable $\Psi$.  Substituting these functions in
Eq.~(\ref{eq:14}) allows the effective control Hamiltonian to now be
written in a compact form
\begin{equation}
   \label{eq:25}
   \CalH =-\sum_{l=1}^{8} a_l\phi_l
-\sum_{l=1}^8 u_l \phi_l+L(u).   
\end{equation}
The optimality condition
\begin{equation}
  \label{eq:15}
\frac{d\CalH}{du}=0
\end{equation}
leads to the following equivalent optimal equations 
\begin{equation}
  \label{eq:44}
\frac{dL}{du_l}=\phi_l,  
\end{equation}
with $l=1$, \dots, $p$. This is a set of algebraic equations that can
be solved to obtain the optimal controls $u_l$ as functions of the
momentum functions $\phi_l$.  The complete set of optimality
conditions are then as follows:
\begin{equation}
  \label{eq:1}
  \begin{aligned}
    \dot \rho&=-[iH, \rho],\\
    \dot \Psi&=-[iH, \Psi],\\
    \rho_0&=\ma{1&0&0\\0&0&0\\0&0&0}, \quad
    \rho_T=\ma{0&0&0\\0&0&0\\0&0&1},\\
\frac{dL}{du_l}&=\phi_l.
  \end{aligned}
\end{equation}

At this point we have arrived at the usual formulation of the
optimality conditions as a two-point boundary-value problem.  As noted
above, in the present situation the numerical challenges associated
with solving this may be avoided by transforming the optimization
problem to an initial value problem for the momentum functions $\phi_l$.
We proceed by first obtaining the time derivative of $\phi_l$:
\begin{eqnarray}
  \label{eq:50}
\dot \phi_l&=&\la \dot \Psi,[X_l,\rho]\ra+\la \Psi,[X_l, \dot\rho]\ra \nonumber\\
&=&\la[-iH,\Psi],[X_l,\rho]\ra+\la \Psi,[X_l, [-iH, \rho]]\ra
\nonumber\\
&=&\la\Psi, [iH, [X_l,\rho]]\ra
-\la \Psi,[X_l, [iH, \rho]]\ra\nonumber\\
&=&\la \Psi,[[iH, X_l],\rho]\ra,
\end{eqnarray}
where the last equality follows from the Jacobi identity~\cite{Sas99}.
For the Hamiltonian given by \Eq{eq:11} we thereby obtain the
following time evolution equation for the momentum functions,
\begin{eqnarray}
  \label{eq:19}
\dot \phi_l &=& \left\la \Psi,\left[\left[
\sum_{j=1}^{8} a_jX_j
+\sum_{j=1}^8 u_j X_j, X_l\right], \rho\right]\right\ra \nonumber\\
&=&\sum_{i=1}^{8} \left(\sum_{j=1}^{8} a_j C_{jl}^i 
+\sum_{j=1}^8 u_j C_{jl}^i\right) \phi_i,
\end{eqnarray}
where we have introduced the structure constants
\begin{equation}
  \label{eq:45}
  [X_i, X_j]=\sum_{k=1}^8 C_{ij}^k X_k.
\end{equation}
Note that the
structure constants are antisymmetric in all the indices, \ie,
\begin{equation}
  \label{eq:3}
 C_{ij}^k=-C_{ji}^k=-C_{kj}^i=-C_{ik}^j,
\end{equation}
for all $i$, $j$, $k\in \{1$, \dots, ${8}\}$. Up to antisymmetry, the
nonzero structure constants are
\begin{equation}
  \label{eq:4}
  \begin{aligned}
 C_{12}^3&=-1, &C_{14}^7&=-2, &C_{15}^6&=1, \\
 C_{24}^6&=-1,  &C_{25}^7&=1, &C_{25}^8&=-\sqrt{3}, \\
 C_{34}^5&=1, &C_{36}^7&=1, &C_{36}^8&=\sqrt{3}.
  \end{aligned}
\end{equation}

\Eq{eq:19} constitutes a set of first order differential equations
that govern the dynamical evolution of the momentum functions
$\phi_l$. When the running cost $L(u)$ in \Eq{eq:13} is chosen as a
quadratic function of the control fields $u$ (as in
Eq.~\eqref{eq:lg_fluence}), these functions $\phi_l$ are linear
combinations of the optimal control fields $u_l$. It is then
straightforward to extract the dynamics of the optimal controls $u$.
The control problem has thereby been reduced to finding an appropriate
initial condition for Eq.~\eqref{eq:19}, a considerably easier task
than solving the two-point boundary value problem of Eq.~\eqref{eq:1}.
  
We note that the functions $\phi_l$ describe the reduced dynamics from the
  Lie-Poisson Reduction Theorem~\cite{Marsden:98}, and \Eq{eq:19} can
also  be derived directly from that Theorem. See Chap. 13 of
  Ref.~\cite{Marsden:98} and Ref.~\cite{Kri:93} for details.

One useful property of $\phi$ is that its norm is a conserved quantity along
the optimal trajectory. This may be shown by recalling that the structure constants
$C_{jl}^i$ are antisymmetric in all the indices, from which we obtain that
\begin{eqnarray*}
&&\frac{d}{dt}\|\phi\|^2 =\frac{d}{dt} \left(\sum_{l=1}^{8} \phi_l^2\right)
=\sum_{l=1}^{8} 2 \phi_l \dot \phi_l\\
&=& 2\sum_{l=1}^{8} \sum_{i=1}^{8}\phi_l\left(\sum_{j=1}^{8} a_j C_{jl}^i 
+\sum_{j=1}^8 u_j C_{jl}^i\right) \phi_i=0.
\end{eqnarray*}
Hence,
\begin{equation}
  \label{eq:32}
\|\phi\|^2= \text{const}.
\end{equation}

\subsection{Numerical algorithm}
We develop a gradient algorithm to find the initial conditions that
optimize the fidelity of the final state.

Consider a given time interval $[0, T]$. Divide it into $N$ equal
intervals $\{[t_k, t_{k+1}]\}_{k=0}^{N-1}$ of length $\Delta
t=t_{k+1}-t_k=T/N$, where $t_0=0$ and $t_{N}=T$.  Note that we must
choose $N$ large enough that the evolution equations for the momentum
functions~\eqref{eq:19} are satisfied.  Within the interval $[t_k,
t_{k+1}]$, assume the control fields $u_l(t)$ takes a constant value
$u_l(k)$ at $t=t_k$.  Define the fidelity of the actually achieved
terminal state $\rho(T)$ as in Eq.~\eqref{eq:16}.  The state transfer
problem amounts to maximizing the fidelity ${F}$ by finding the
optimal control pulses $u_l(k)$.

From \Eq{eq:44}, we know that the control fields $u_l(k)$ are
determined once the momentum functions $\phi_l(k)$ are known.
Furthermore, the $\phi_l(k)$ are obtained by solving \Eq{eq:19} with
the initial condition $\phi(0)$ (the vector with components
$\phi_l(0)$). Therefore, to maximize the fidelity, we just need to
find an appropriate vector $\phi(0)$. The advantage of optimizing over
$\phi(0)$ instead of over $u_l$ is that $\phi(0)$ is a vector with
dimension 8, whereas $u_l$ has dimension $N$, which is usually a much
larger number.

The gradient of the fidelity ${F}$ with respect to $\phi_l(0)$,
${d{F}}/{d\phi_l(0)}$, can be derived explicitly: details are
presented in Appendix A.  With this gradient in hand we can then
formulate a gradient algorithm to determine the optimal initial
condition $\phi^*(0)$.  The other components of this gradient
algorithm are solution of the coupled first order equations,
Eq.~(\ref{eq:19}) to obtain $\phi(t)$ and solution of
Eq.~(\ref{eq:44}) to obtain the physical control fields $u_l(t)$ from
the $\phi(t)$.  The full algorithm is then constructed as follows.

\begin{algorithm}\mbox{}
 \begin{enumerate}
\item Choose an initial guess for $\phi^0(0)$;
\item  At the $j$-th step, solve the differential equation \Eq{eq:19}
  with the initial condition $\phi^{j}(0)$ to get $\phi_l(k)$;
\item Solve the algebraic equation \Eq{eq:44} to get the optimal
  controls $u_l(k)$ as functions of $\phi_l(k)$;
\item Follow the procedure in Appendix A to derive
  $\nabla_{\phi^j(0)} {F}$;
\item Let $\phi^{j+1}(0)=\phi^{j}(0)+\epsilon\nabla_{\phi^j(0)} {F} $,
  where $\epsilon$ is a small positive number;
\item Repeat Steps (2)--(5) until a desired fidelity is reached.
\end{enumerate}
\end{algorithm}

Note that Khaneja {\it et al}~\cite{Khaneja:05} developed the gradient
ascent pulse engineering (GRAPE) algorithm to solve a similar problem.
The difference between GRAPE and our algorithm is that GRAPE solves
for the control pulses directly, whereas our algorithm optimizes over
the initial condition of a differential equation.

We will apply our algorithm to two physical systems, the triple
quantum dot system discussed in Ref.~\cite{Chen:11} and the ionized
donor chain discussed in Ref.~\cite{Greentree:04}.  For the ionized
donor chain we further show that the optimized control fields can also
provide a high degree of spin state transfer when the shuttled
electron is coupled to the donor nuclei by the hyperfine interaction.

\section{Triple Quantum Dot\label{sec:lg_chen}}
We now investigate electron shuttling for the triple quantum dot
system discussed in Ref.~\cite{Chen:11}.  In this system, an electron
beginning in the left dot of a three lateral quantum dot system is
moved to the right dot.  The relative energies of the left and right
dots are controlled by external gate voltages.  The Hamiltonian is
given by
\begin{equation}
  \label{eq:9}
H=\ma{\mu_L(t) &J_1&0\\J_1&0&J_2\\0&J_2&\mu_R(t) },
\end{equation}
where the control fields are the on-site energies $\mu_L$ and $\mu_R$,
and $J_1$ and $J_2$ are the fixed coupling constants between nearest
neighboring dots. Using the basis in \Eq{eq:2}, we can rewrite the
Hamiltonian \eqref{eq:9} as
\begin{equation}
  \label{eq:28}
   iH=J_1X_1+J_2X_2+\frac{\mu_L}{2}X_7
 +\frac{\mu_L-2\mu_R}{2\sqrt{3}}X_8
+\frac{\mu_L+\mu_R}{3}i I_3
\end{equation}
We consider the minimum energy cost function (see
Sec.~\ref{sec:lg_cc}):
\begin{equation*}
\min \frac12\int_0^T \left(\mu_L^2(\tau) + \mu_R^2(\tau)
\right) d\tau.
\end{equation*}
The parameters in the Hamiltonian of Eq.~\eqref{eq:11} are
\begin{equation*}
  a_1=J_1, \quad a_2=J_2, \quad u_7=\frac{\mu_L}2,\quad 
u_8=\frac{\mu_L-2\mu_R}{2\sqrt{3}}.
\end{equation*}
Hence
\begin{equation*}
  \mu_L=2u_7,\quad \mu_R=u_7-\sqrt{3} u_8,
\end{equation*}
and the running cost is
\begin{equation*}
  L(u)=\frac{\mu_L^2}2+\frac{\mu_R^2}2
=2u_7^2+\frac{(u_7-\sqrt{3} u_8)^2}2.
\end{equation*}
The optimality condition~\eqref{eq:44} becomes
\begin{equation*}
  \phi_7=5u_7-\sqrt{3}u_8, \quad \phi_8=3u_8-\sqrt{3}u_7,
\end{equation*}
which yields 
\begin{equation*}
  \label{eq:22}
 u_7=\frac{\sqrt{3}\phi_7+\phi_8}{4\sqrt{3}}, \quad
\quad u_8=\frac{\sqrt{3}\phi_7+5\phi_8}{12},
\end{equation*}
and hence
\begin{equation*}
 \mu_L=\frac{\sqrt{3}\phi_7+\phi_8}{2\sqrt{3}}, \quad
\quad \mu_R=-\frac{\phi_8}{\sqrt{3}}.
\end{equation*}
The dynamics of the momentum functions $\phi$ are obtained from Eq.~\eqref{eq:19} as
\begin{equation}
\label{eq:29}
\begin{aligned}
\dot \phi_1&=J_2 \phi_3-\phi_4 \phi_7/2- \sqrt{3}/6 \phi_4 \phi_8 \\
\dot \phi_2&=-J_1 \phi_3-\phi_5 \phi_8/\sqrt{3}  \\
\dot \phi_3&=-J_2 \phi_1+J_1 \phi_2+\phi_6 \phi_7/2+\sqrt{3}/2 \phi_6 \phi_8 \\
\dot \phi_4&=\phi_1 \phi_7/2+\sqrt{3}/6 \phi_1 \phi_8-J_2 \phi_6-2 J_1 \phi_7 \\
\dot \phi_5&=\phi_2\phi_8/\sqrt{3}+J_1 \phi_6+J_2 \phi_7-\sqrt{3}J_2 \phi_8 \\
\dot \phi_6&=-\phi_3 \phi_7/2- \sqrt{3}/2\phi_3 \phi_8+J_2 \phi_4-J_1 \phi_5 \\
\dot \phi_7&=2 J_1 \phi_4-J_2 \phi_5 \\
\dot \phi_8&=\sqrt{3}J_2 \phi_5 .
\end{aligned}
\end{equation}
For the derivation of the gradient of the fidelity and an explicit expression for this system, see Appendices A and B.

\begin{figure}[tb]
\begin{center}
\begin{tabular}{cccc}
\includegraphics[width=0.4\hsize]{\figurelib/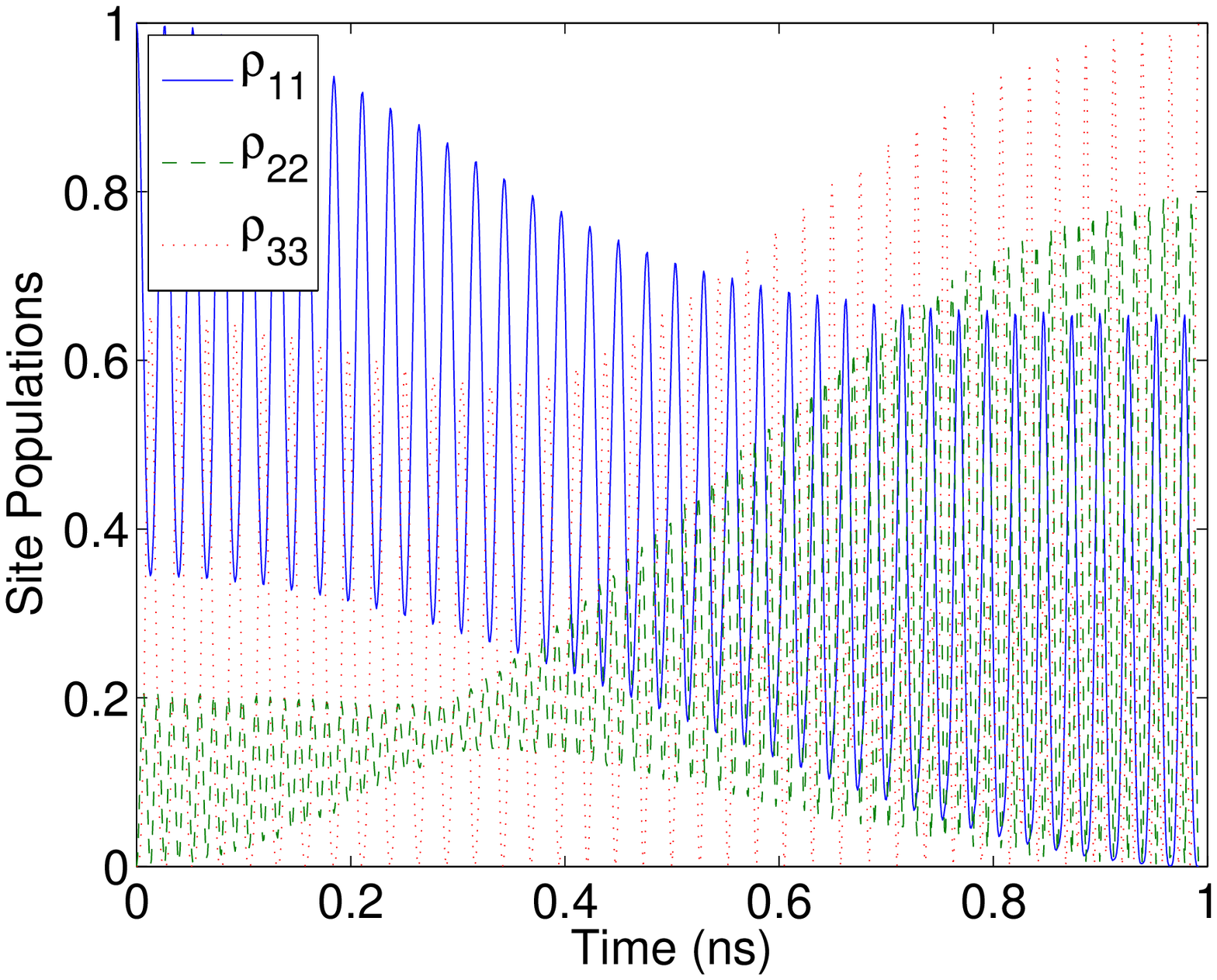}
& (A)
&\includegraphics[width=0.4\hsize]{\figurelib/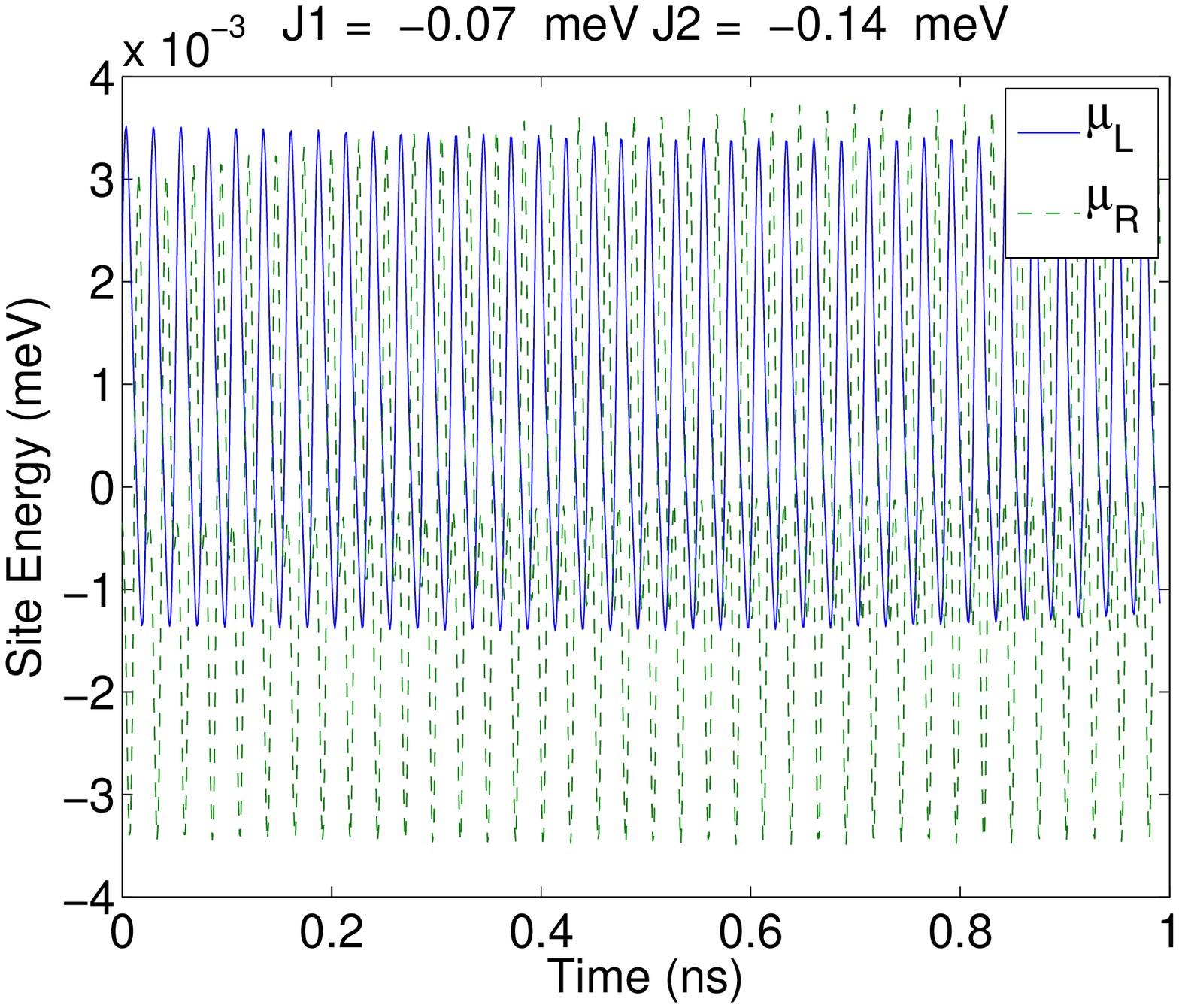}
& (B)\\
\includegraphics[width=0.4\hsize]{\figurelib/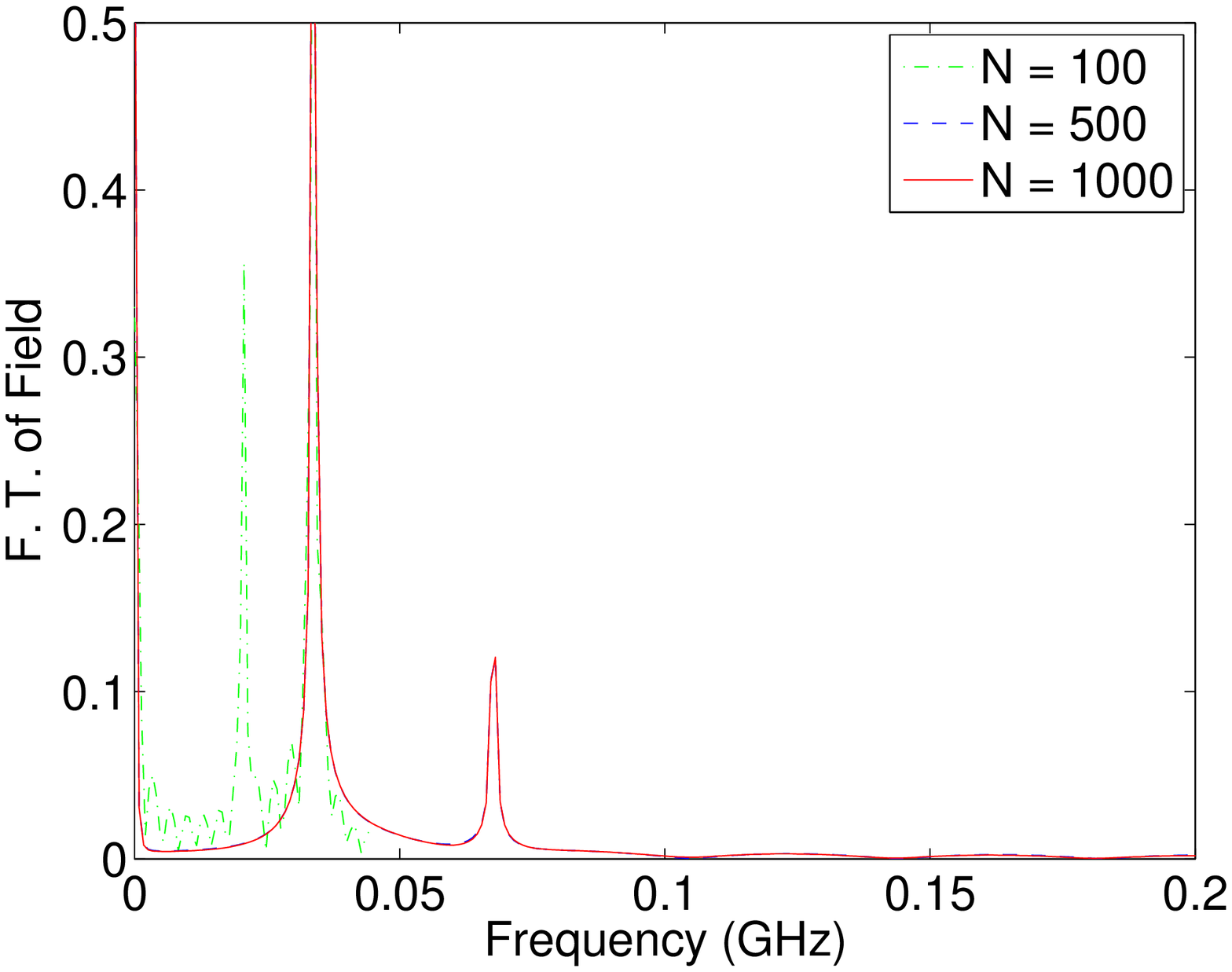}
& (C)
&\includegraphics[width=0.4\hsize]{\figurelib/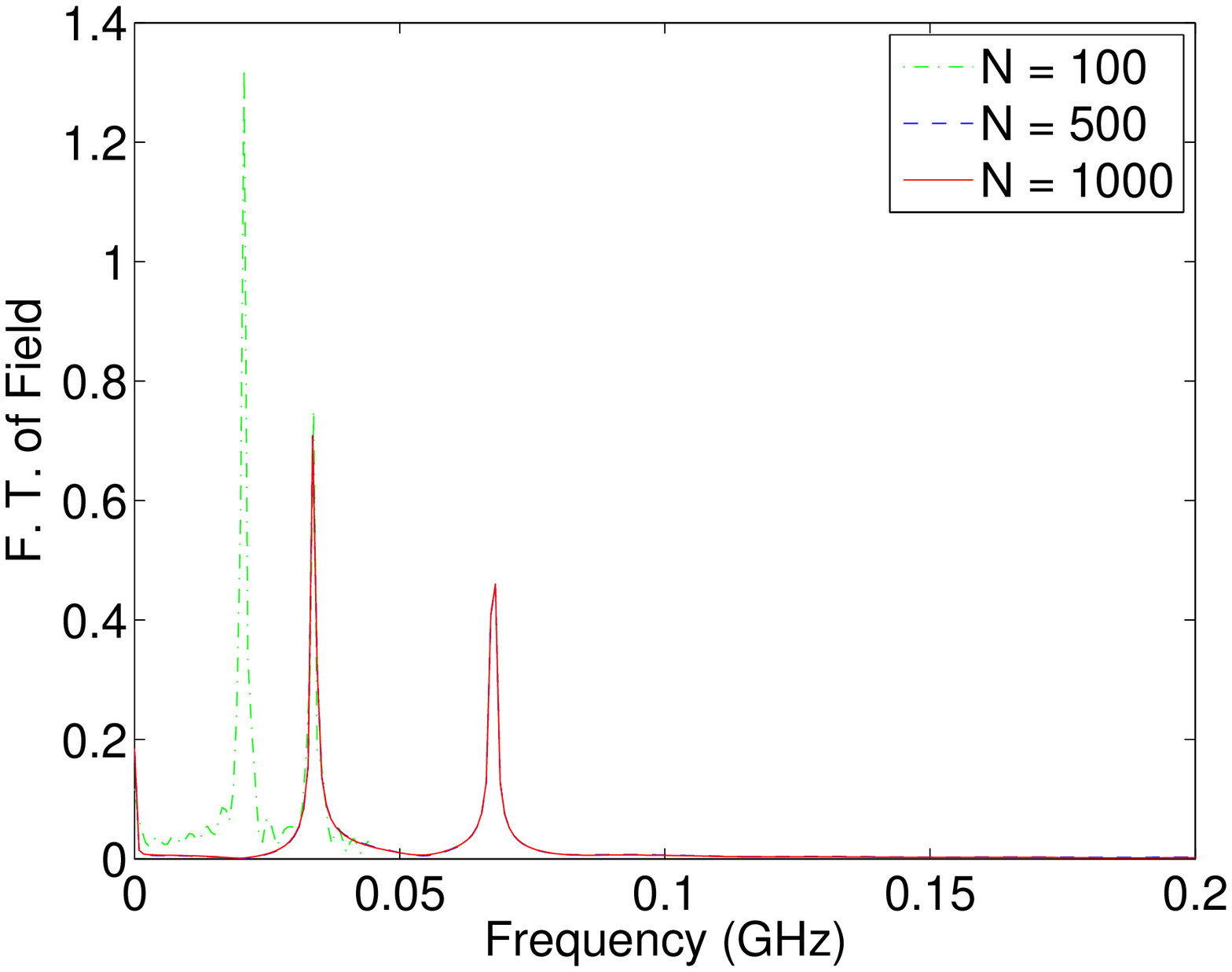}
& (D)\\
\end{tabular}
\end{center}
\caption{(color online) Time dependence of site populations for 
  electron shuttling across a triple quantum dot when acted on by time
  dependent voltages optimized to achieve minimal heating (i.e.,
  minimal pulse energy). (A) Quantum dot populations for sites 1, 2,
  and 3, as a function of time. Blue solid line: $\rho_{11}$; Green
  dashed line: $\rho_{22}$; Red dotted line: $\rho_{33}$.  $J_1$ and
  $J_2$ were set to -0.07 and -0.14 meV, respectively.  (B) Optimal
  control voltages.  Blue solid: $\mu_L$; Green dashed: $\mu_R$.  The
  pulses were determined here for $N=1000$ segments.  (C) and (D) The
  Fourier transform of $\mu _L$ and $\mu _R$, respectively, at
  different numbers of segments.  We note that the form of the pulses
  converge at $N=500$, after which pulses and site populations are
  indistinguishable from the corresponding values obtained with
  $N=1000$.  }
 \label{fig:2}
\end{figure}

The optimized pulses for transfer with $J_1$ set to -0.07 meV and $J_2$ set to -0.14 meV are given in Fig.~\ref{fig:2}.
The transfer time was taken to be 1 ns.
The algorithm converges at $N=500$ slices of the time interval.
The correlation between neighboring time steps can be seen as the optimized pulses are dominated by a small number of frequency components.

\section{Ionized Donor Chain\label{sec:lg_greentree}}
In this section we apply our control algorithm to the ionized donor
chain studied in Ref.~\cite{Greentree:04}.  The system consists of
three singly ionized phosphorus donors in silicon, and one electron
shared in the system.  The electron begins on the first phosphorus,
site 1, and the pulses are designed to move this electron to site 3.
The Hamiltonian is given by:
\begin{equation}
  \label{eq:33}
  H=\ma{0&-\Omega_{12}(t) &0\\-\Omega_{12}(t) &\Delta&-\Omega_{23}(t) \\
0&-\Omega_{23}(t) &0}.
\end{equation}
Here the control terms are $\Omega_{12}$ and $\Omega_{23}$, which are
the coherent tunneling rate between adjacent dopants. Under the basis
in \Eq{eq:2}, this Hamiltonian can be written as
\begin{equation}
  \label{eq:34}
  iH=-\Omega_{12} X_1-\Omega_{23}X_2-\frac{\Delta}2 X_7
+\frac{\Delta}{2\sqrt{3}}X_8+\frac{\Delta}3 i I_3.
\end{equation}
We can drop the term $\frac{\Delta}3 I_3$ as it commutes with all the
other terms and thus contributes only a global phase. Consider the
following minimum energy cost function
\begin{equation*}
\min \frac12\int_0^T \left(\Omega_{12}^2(\tau) +\Omega_{23}^2(\tau)
\right) d\tau
\end{equation*}
with the initial and terminal states given in Eqs.~\eqref{eq:6}-\eqref{eq:7}.
Following the procedure in Sec.~\ref{sec:oc}, we find that the parameters
in the Hamiltonian of Eq.~\eqref{eq:11} are
\begin{equation*}
  \label{eq:8}
  u_1=-\Omega_{12}, \quad u_2=-\Omega_{23}, \quad
a_7=-\frac{\Delta}2, \quad a_8=\frac{\Delta}{2\sqrt{3}},
\end{equation*}
and the running cost is
\begin{equation*}
  L(u)=\frac{u_1^2}2+\frac{ u_2^2}2.
\end{equation*}
The optimality condition~\eqref{eq:44} yields 
\begin{equation*}
  \phi_1=u_1,\quad \phi_2=u_2,
\end{equation*}
and hence the optimal controls are given by
\begin{equation}
  \label{eq:20}
  \Omega_{12}=-\phi_1,\quad \Omega_{23}=-\phi_2.
\end{equation}
For the Hamiltonian of Eq.~\eqref{eq:34}, the dynamics of $\phi$ in
\Eq{eq:19} becomes
\begin{equation}
  \label{eq:63}
\dot \phi_l =\sum_{i=1}^{8} \left(\phi_1 C_{1l}^i+\phi_2 C_{2l}^i
-\frac{\Delta}2 C_{7l}^i +
\frac{\Delta}{2\sqrt{3}} C_{8l}^i \right) \phi_i,
\end{equation}
Substituting the values of structure constants $C_{ij}^k$ in \Eq{eq:4}
into \Eq{eq:63} yields the complete dynamics of $\phi$:
\begin{equation}
  \label{eq:53}
  \begin{aligned}
\dot \phi_1&=\phi_2 \phi_3+\Delta \phi_4\\
\dot \phi_2&=-\phi_1 \phi_3-\Delta \phi_5\\
\dot \phi_3&=0 \\
\dot \phi_4&=-\Delta \phi_1-\phi_2 \phi_6-2 \phi_1 \phi_7\\
\dot \phi_5&=\Delta \phi_2+\phi_1 \phi_6+\phi_2 \phi_7-\sqrt{3} \phi_2\phi_8\\
\dot \phi_6&=\phi_2 \phi_4-\phi_1 \phi_5\\
\dot \phi_7&=2 \phi_1 \phi_4-\phi_2 \phi_5\\
\dot \phi_8&=\sqrt{3} \phi_2 \phi_5.
  \end{aligned}
\end{equation}
The required matrices for the gradient algorithm for this systems are given in Appendix C.

\begin{figure}[tb]
\begin{center}
\begin{tabular}{cc}
\includegraphics[width=0.8\hsize]{\figurelib/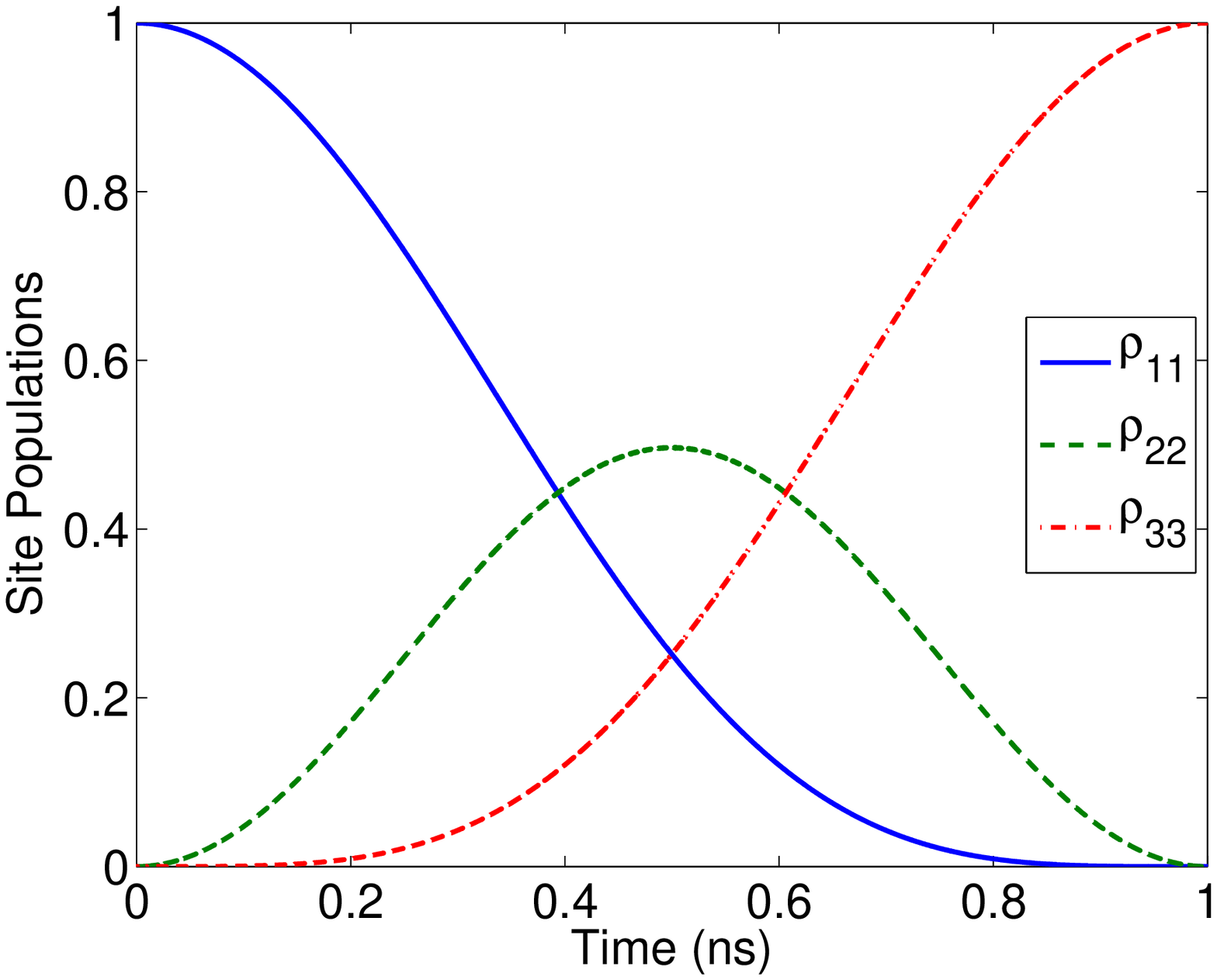} 
& (A)\\
\includegraphics[width=0.8\hsize]{\figurelib/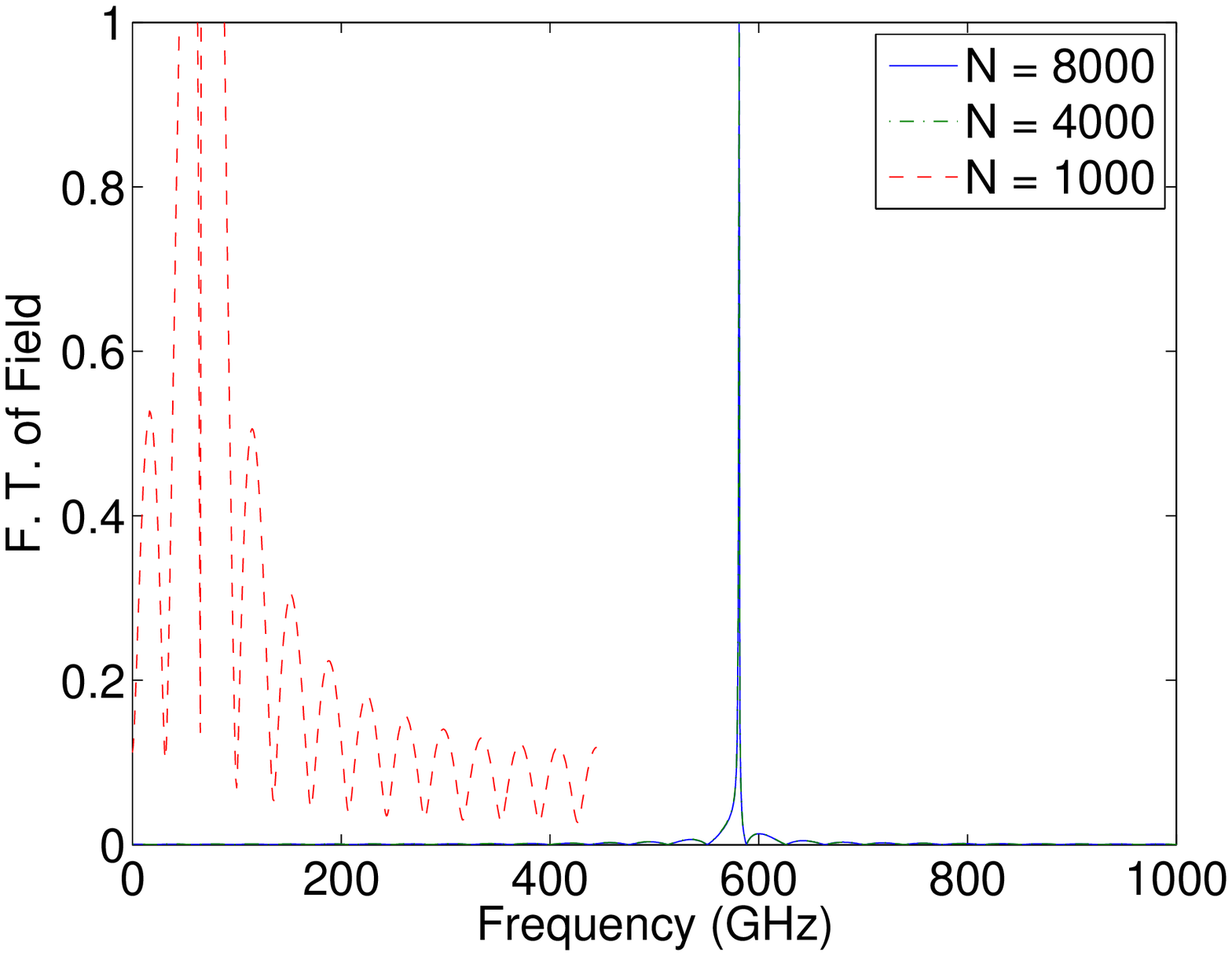}
& (B)
\end{tabular}
\end{center}
\caption{(color online) Time dependence of site populations for
  electron shuttling across a chain of three singly ionized phosphorus
  ions, when acted on by time dependent voltages optimized to achieve
  minimal heating (i.e., minimal pulse energy). (A) Electron
  populations on donor sites 1, 2, and 3, as a function of time.
  Blue solid: $\rho_{11}$; Green dashed: $\rho_{22}$; Red dash-dot:
  $\rho_{33}$.  The value of $\Delta$ was set to 2.7 meV.  (B) Fourier
  transform of the optimal control pulse $\Omega _{12}$.  The second
  control pulse $\Omega _{23}$ has the same frequency components as
  $\Omega _{12}$ with a phase difference of -1.719 rad.  The
  time-domain pulses oscillate with a very high frequency,
  corresponding to the 2.7 meV value of $\Delta$, and so are not shown
  here.  The optimal pulses are found converge at $N=8000$ segments.}
 \label{fig:1}
\end{figure}

\begin{figure}[tb]
\begin{center}
\begin{tabular}{cc}
\includegraphics[width=0.8\hsize]{\figurelib/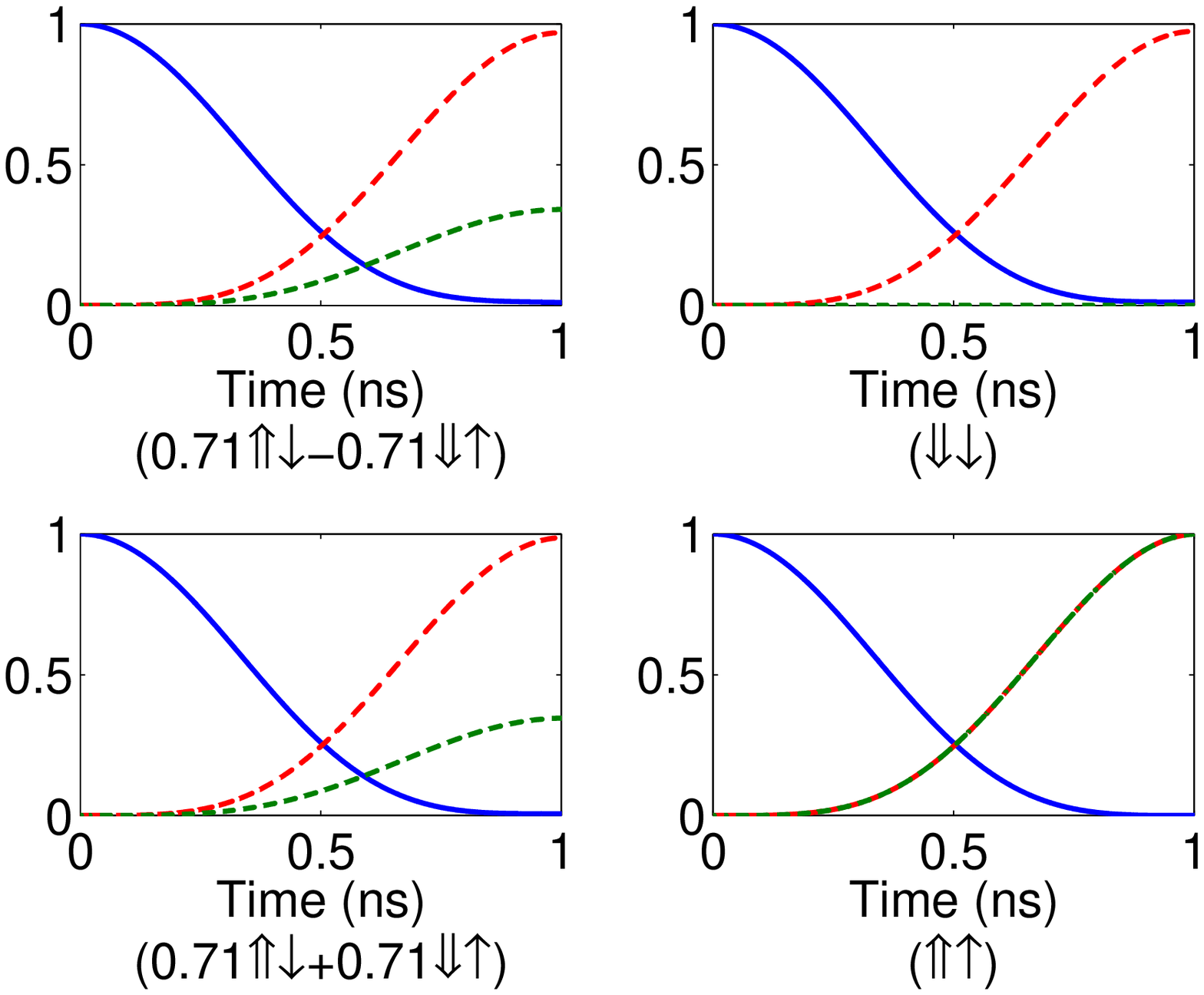} 
& (A)\\
\includegraphics[width=0.8\hsize]{\figurelib/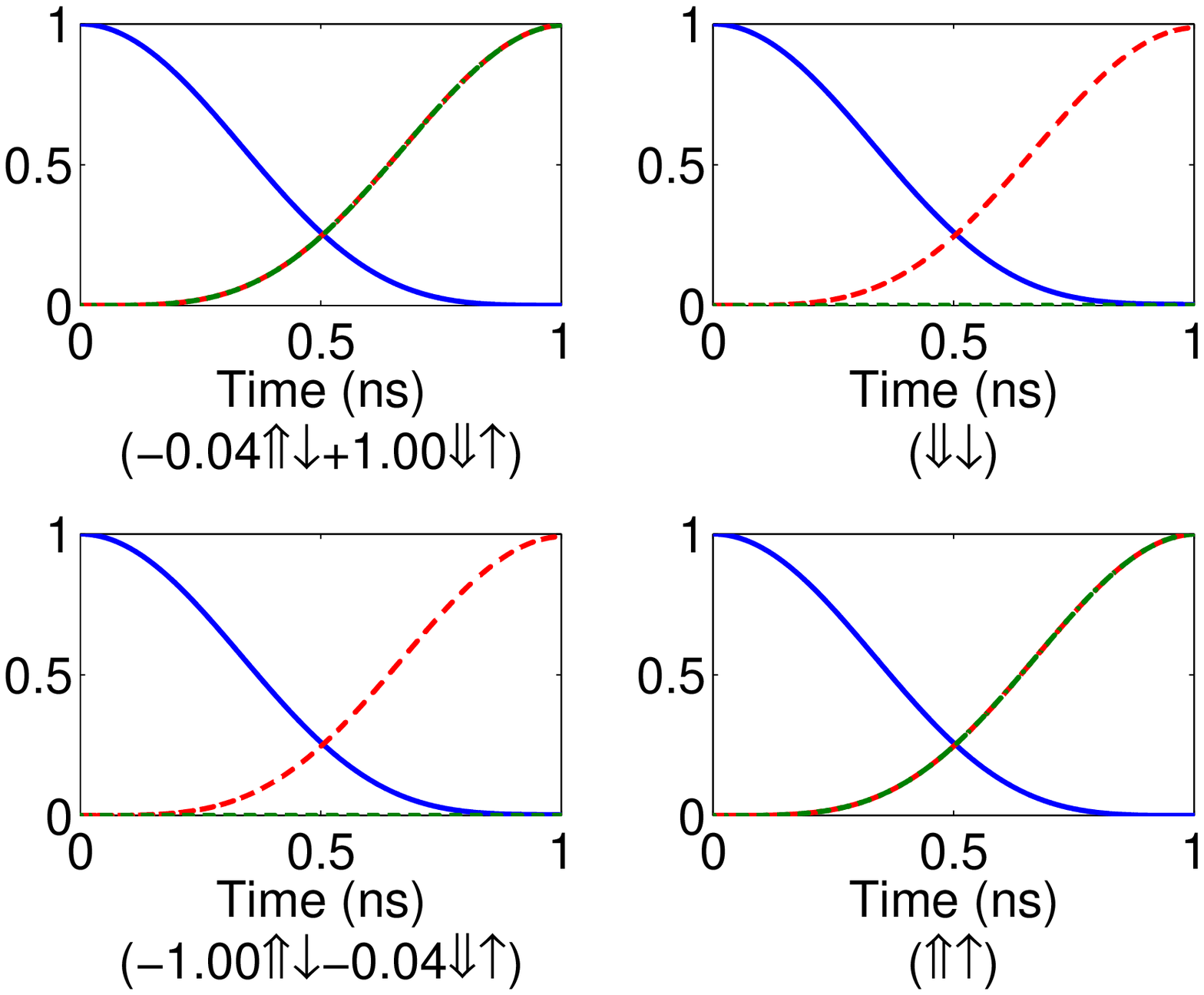}
& (B)
\end{tabular}
\end{center}
 \caption{(color online) (A) 
   Time dependence of site populations on for electron shuttling
   across a chain of three single ionized phosphorus atoms when the
   spatial shuttling Hamiltonian is supplemented by the spin
   Hamiltonian, \eqref{eq:lg_spinham} at zero magnetic field.  We show
   the transfer of all four hyperfine eigenstates accessible to the
   electron on site 1, under the pulses optimized solely for spatial
   shuttling in Fig.~\ref{fig:1}.  The solid blue and dashed red lines
   show the population on sites 1 and 3, respectively, as before.  The
   dashed green line shows measure D (Eq.~(\ref{distmeas})) of the
   hyperfine state transfer at site 3.  (B) Same as (A), but in the
   presence of a finite magnetic field (500 G).  The figures show that
   hyperfine states which align the nuclear spin with the magnetic
   field can be robustly transferred, independent of the magnetic
   field value, while transfer of the spin flipped states is
   energetically forbidden. See text for detailed explanation.}
 \label{fig:3}
\end{figure}

In Fig.~\ref{fig:1}, the optimized pulses are shown for $\Delta = 2.7$
meV and a transfer time of 1 ns.  This choice of parameters is consistent with the values
calculated using tight binding theory~\cite{Rahman:10}.  In
Fig.~\ref{fig:1}(A), the populations of each site are shown as a
function of time, the population is fully transferred from the first
to the third site.  The maximum magnitude of the pulses is on the
order of magnitude of $10^{-4}$ meV.  Using the guideline for
adiabatic transfer in Ref.~\cite{Greentree:04}, $3.75 \approx \Omega
_{max} t_{max} / \pi$, this pulse magnitude would require a transfer
time of 2.5 ns, or conversely the transfer time of 1 ns would require
a pulse 2.5 times larger.

Since one of the main qubits of interest for solid-state quantum logic
is phosphorus-doped silicon, where quantum information may be stored
in either or both the spin of the electrons and
nuclei~\cite{Kane_Proposal,Skinner_Proposal,Morton_Nature_Review} we
have also investigated the performance of these optimal shuttling
pulses in transmitting a hybrid electron-nuclear hyperfine spin state
together with the spatial transfer of the electron.  Here we assess
the robustness of this procedure with respect to the spin states.

The hyperfine interaction was modeled as an on-site interaction of the
electron spin ($\sigma _e$) with the spin of the nucleus at each site
$i$ ($\sigma _{N_i}$).  To this we add the Zeeman interaction of each
spin with the magnetic field $B$, to obtain the spin Hamiltonian
\begin{equation}
\label{eq:lg_spinham}
H_{spin} = B\gamma _e \sigma _e^z + \sum _i A\sigma _e \cdot 
\sigma _{N_i} \bra{i}\ket{i} -B\gamma _N \sigma _{N_i}^z,
\end{equation}
where $A$ is the hyperfine constant and $\gamma _e$ and $\gamma _{N}$
are the electron and nuclear gyromagnetic ratios.  Note that we have
chosen the sign convention in which $\gamma_e$ is positive.  The
eigenstates of the spin Hamiltonian (Eq.~\eqref{eq:lg_spinham}) can be
used to store quantum information.  These states consist of the
electron-nuclear spin aligned states $\ket{\Uparrow \uparrow}$ and
$\ket{\Downarrow \downarrow}$, and linear combinations of the
anti-aligned states $\ket{\Uparrow \downarrow}$ and $\ket{\Downarrow
  \uparrow}$, where the double arrows represent the electron spin and
the single arrows represent the nuclear spin.  As the magnetic field
is increased, the eigenstates are dominated by one of the anti-aligned
states, and at zero magnetic field the eigenstates are an equal
superposition.  Combined with the spatial Hamiltonian of
Eq.~\eqref{eq:33}, the entire Hamiltonian then given by
\begin{equation}
\begin{aligned}
H =& -\Omega _{12}(t) \left(\bra{1}\ket{2}+\bra{2}\ket{1}\right) \\
& -\Omega _{23}(t) \left(\bra{2}\ket{3}+\bra{3}\ket{2}\right) \\
& +\Delta \bra{2}\ket{2} + B \gamma _e \sigma _e^z -
B\gamma _N\left(\sigma_{N_1}^z+ \sigma _{N_2}^z+ \sigma _{N_3}^z\right)\\
&+ A\left( \sigma _e \bra{1}\ket{1} \cdot \sigma _{N_1}
+ \sigma _e \bra{2}\ket{2} \cdot \sigma _{N_2}
+ \sigma _e \bra{3}\ket{3} \cdot \sigma _{N_3}
\right).
\end{aligned}
\end{equation}

For the phosphorus donor system, a hyperfine interaction with a
splitting of $A=117.5$ MHz was used~\cite{Feher_ESR1}.  Results are
shown in Fig.~\ref{fig:3} for no external magnetic field (A) and for a
field of 500 G (B).  The spins of the nuclei at sites 2 and 3 are
initialized into the $\uparrow$ state, while on site 1 the
electron-nuclear system is initialized into one of four hyperfine
eigenstates (each panel of Fig.~\ref{fig:3} represents starting in a
different hyperfine eigenstate; the coefficients of each eigenstate
are shown under the figure).  The distance measure ($D$) shown in
Fig.~\ref{fig:3} is a measure of the fidelity of transfer of this
hyperfine state,
\begin{equation}
D=1-\vert \vert \rho _{T}-\rho _{hf}\vert \vert _2.
\label{distmeas}
\end{equation}
Here $\rho _{hf}$ is the density matrix for one of the hyperfine pure
states (the spin-aligned states or the anti-aligned linear combinations)
, and $\rho _{T}$ is the reduced density matrix of the site 3
nuclear and electron spin at the end of the spatial transfer.  The
norm used in Eq.~(\ref{distmeas}) is the induced 2-norm of the
difference matrix, also known as the spectral norm, which is the
maximum singular value of the matrix~\cite{Callier:91}.  We have also
calculated the fidelity~\cite{Nielsen:00,Fuchs:96}, the trace
distance~\cite{Nielsen:00}, and the Frobenius norm of the difference
matrix~\cite{Callier:91}.  While all norms show a similar picture
regarding which states are transferred, the measure $D$ has the
pictorial advantage of following the population on the third site when
full transfer is occurring, as well as remaining zero when the fidelity
is zero, unlike the Frobenius norm.  

At all magnetic fields, the $\vert \Uparrow \uparrow \rangle$ state
can be transferred completely from site 1 to site 3, because with all
of the nuclear spins up the electrons remain in the hyperfine
eigenstate, no matter which spatial site it is on.  Conversely, the
hyperfine state $\vert \Downarrow \downarrow \rangle$ cannot be
transferred at any magnetic field value, because transfer of this spin
state requires flipping the spins of the nuclei on sites 2 and 3,
which is not allowed energetically.  The corresponding spatial
fidelities of the state transfer are given in Table~\ref{tab:spatfid}
for all possible initial spin eigenstates.  It is evident that the
transfer fidelity for the spatial degrees of freedom are only slightly
affected by the spin interactions.  Additional calculations have shown
that for larger hyperfine constants the spatial transfer of the
electron can be reduced significantly in the presence of the hyperfine
interaction.  For the linear combinations of spin states, $\vert
\Uparrow \downarrow \rangle$ and $\vert \Downarrow \uparrow \rangle$,
partial spin transfers can be accomplished corresponding to the
contribution from the component which has the nuclear spin up.  This
can be understood because as the magnetic field is turned on
(Fig.~\ref{fig:3}(B)) and the relative magnitude of the two components
in the linear combination becomes asymmetric, the transfer of the
component which becomes primarily nuclear spin up can be achieved
while transferring the component which becomes primarily nuclear spin
down cannot.  It should be noted that even in the case of a 500 G
magnetic field, two hyperfine states can nevertheless be transferred
spatially with high fidelity, suggesting their potential use as a
mobile qubit.

\begin{table}[htp!]
\caption{The spatial fidelity for transfer starting from the left 
dopant in a given hyperfine eigenstate.  
The first and third columns give the eigenstate at $B = 0 G$ and $B = 500 G$, respectively.
The second and fourth columns give the corresponding spatial fidelity.\label{tab:spatfid}}
\begin{ruledtabular}
\begin{tabular}{cdcd}
\multicolumn{2}{c}{$B = 0 G$}&
\multicolumn{2}{c}{$B = 500 G$}\\
\cline{1-2}\cline{3-4}
$0.71\Uparrow \downarrow - 0.71\Downarrow \uparrow$&0.9691&
$-0.04\Uparrow \downarrow + 1.00\Downarrow \uparrow$&0.9970\\
$\Downarrow \downarrow$&0.9740&$\Downarrow \downarrow$&0.9873\\
$0.71\Uparrow \downarrow + 0.71\Downarrow \uparrow$&0.9870&
$-1.00\Uparrow \downarrow - 0.04\Downarrow \uparrow$&0.9913\\
$\Uparrow \uparrow$&1.00&$\Uparrow \uparrow$&1.00
\end{tabular}
\end{ruledtabular}
\end{table}

\section{Conclusion} 
In this paper we have formulated the general problem of solid-state
electron shuttling as a state transfer problem in optimal control
theory. We derived the underlying dynamical equations that govern the
time evolution of optimal control fields.  Use of a momentum function
was shown to lead to an effective algorithm with a small number of
optimizing variables that requires numerical solution of an initial
value problem rather than a two-point boundary problem.  We
demonstrated the efficacy of our algorithm with application to two
physical examples.

First, we determined the control pulses for state transfer between
left and right quantum dots in a triple quantum dot system.  Since
hyperfine interactions in lateral quantum dots can be small, such
spatial transfer allows the ability to transmit quantum information
and possibly use this ability to couple qubits.  Second, we applied
the optimal control approach to the system of shuttling of an electron
along an ionized phosphorus donor chain in silicon.  We again
determined optimal control pulses for spatial transfer, finding a
significant reduction in time and energy of the optimal pulses
compared to those required by adiabatic protocols.  For the shuttling
across donor chains we also expanded the Hamiltonian to include
magnetic interactions of the electron spin with the donor nuclear
spins and of both electron and nuclear spins with external magnetic
fields and then investigated the robustness of transfer of the
hyperfine spin state states under these optimal pulses to variations
in the magnetic field strength.  For a magnetic field strength of 500
G (0.05 T), we find that two hyperfine states of the electron-nucleus
on a given donor can be transferred across the chain to a distant
donor with high fidelity.  As the external field is decreased to zero,
however, only one of four hyperfine states can be spatially
transferred with high fidelity.  Therefore, in order to transfer spin
quantum information in a donor chain within a low-field environment,
it will be necessary to design control pulses which are optimized for
both spin and spatial dynamics.  This will be addressed in a future
publication.

\begin{acknowledgments}
  JZ thanks the financial support from the Innovation Program of
  Shanghai Municipal Education Commission under Grant No. 11ZZ20,
  Shanghai Pujiang Program under Grant No. 11PJ1405800, NSFC under
  Grant No. 61174086, Project-sponsored by SRF for ROCS SEM, State Key
  Laboratory of Precision Spectroscopy (ECNU), and State Key Lab of
  Advanced Optical Communication Systems and Networks (SJTU), China.
  LG and KBW thank NSA (Grant No. MOD713106A) for financial support.
  XD thanks the University of California-Berkeley College of Chemistry
  for a Summer Research Stipend.

\end{acknowledgments}

\appendix
\section{Derivation of $\frac{d{F}}{d\phi_l(0)}$}
\label{App:A}
In this appendix, we derive the gradient of the fidelity ${F}$ with
respect to $\phi(0)$.  This gradient of performance with respect to
initial conditions of the momentum functions is required for step 4 of
the gradient algorithm and is the key component of the algorithm to
find the optimal control fields $u_l(t)$.  By the chain rule, we have
\begin{eqnarray}
  \label{eq:55}
&&\frac{d{F}}{d\phi_l(0)}=
\sum_{k=0}^{N-1}\sum_{m\in \overline{M}} 
\frac{d{F}}{du_m(k)}\frac{d u_m(k)}{d\phi_l(0)}\nonumber\\
&=&\sum_{k=0}^{N-1} \sum_{m=1}^p  \sum_{s=1}^{8}
\frac{d{F}}{du_m(k)}\frac{d u_m(k)}{d\phi_s(k)}
\frac{d \phi_s(k)}{d\phi_l(0)},
\end{eqnarray}
where $\overline{M}$ is an index set of all the control fields.
We thus need to derive the three differentials in each term on the right hand side
of \Eq{eq:55}. 

\noindent {\it 1.} We first consider the second term, $\frac{d u_m(k)}{d\phi_s(k)}$. From
\Eq{eq:44}, we obtain
\begin{equation}
  \label{eq:57}
 \frac{d}{d\phi_s} \left(\frac{dL}{du_m}\right)
=\frac{d}{d\phi_s} \phi_m =\delta_{ms}.
\end{equation}
Because the running cost $L$ is defined as a function of only the
control fields $u$ as in \Eq{eq:13}, $\frac{dL}{du_m}$ is also a
function of $u$ only. Therefore, we can obtain $\frac{d
  u_m(k)}{d\phi_s(k)}$ by solving the algebraic equation \Eq{eq:57}.
For example, when $L$ is taken as a quadratic function
$L=\frac{1}{2}(u_1^2+u_2^2)$, it is straightforward to show that
$\frac{du_m}{d\phi_s}=\delta_{ms}$.

\noindent {\it 2.} Next we consider the third term,
$\frac{d\phi_s(k)}{d\phi_l(0)}$, {\it i.e.} the derivative
of the momentum functions with respect to their initial conditions. These derivations may be 
obtained from~\Eq{eq:19}.  
We rewrite \Eq{eq:19} as a vector differential equation 
\begin{equation}
  \label{eq:47}
  \dot \phi={S}(\phi),
\end{equation}
with $\phi=\ma{\phi_1&\dots&\phi_{8}}$ and where we have used the
relation between the control functions $u$ and the momentum functions
$\phi$ given by \Eq{eq:44} to write the right hand side as a function
of $\phi$ alone, {\it i.e.} $S(\phi)$.  Note that the form of
$S(\phi)$ will depend on the form of the cost function $L$.
Differentiating both sides of \Eq{eq:47} with respect to $\phi(0)$, we
now obtain
\begin{eqnarray}
  \label{eq:5}
  \frac{d}{d \phi(0)}\dot \phi= \frac{d}{d \phi(0)} {S}(\phi)
= D{S}(\phi) \frac{d \phi}{d\phi(0)},
\end{eqnarray}
where the Jacobian matrix $D{S}(\phi)$ is given by
\begin{equation}
  \label{eq:59}
D{S}(\phi)=\ma{ \ppfrac{{S}_1}{\phi_1}&\cdots&\ppfrac{{S}_1}{\phi_{8}}\\
\vdots&&\vdots\\
\ppfrac{{S}_{8}}{\phi_1}&\cdots&\ppfrac{{S}_{8}}{\phi_{8}}},
\end{equation}
and
\begin{equation}
  \label{eq:67}
\frac{d\phi}{d\phi(0)}=\ma{ \ppfrac{\phi_1}{\phi_1(0)}&\cdots
&\ppfrac{\phi_1}{\phi_{8}(0)}\\
\vdots&&\vdots\\
\ppfrac{\phi_{8}}{\phi_1(0)}&\cdots&\ppfrac{\phi_{8}}{\phi_{8}(0)}}.  
\end{equation}
From Proposition 6.1 of Chapter 1 in Ref.~\cite{Taylor:96}, we have
\begin{equation}
  \label{eq:10}
\frac{d}{d \phi(0)}\dot \phi= \frac{d}{dt} \frac{d\phi}{d \phi(0)},
\end{equation}
that is, it is legitimate to change the order of the differentials
with respect to $t$ and $\phi(0)$. Combining \Eq{eq:5} and \Eq{eq:10},
we then arrive at the following differential equation that is
satisfied by $\frac{d\phi}{d \phi(0)}$:
\begin{eqnarray}
  \label{eq:58}
\frac{d}{dt} \frac{d\phi}{d \phi(0)}= D{S}(\phi) \frac{d \phi}{d\phi(0)},
\end{eqnarray}
with initial condition 
\begin{equation}
  \label{eq:60}
 \left.\frac{d\phi}{d\phi(0)}\right|_{t=0}=I.
\end{equation}
Solving this differential equation \Eq{eq:58}, with initial condition
\Eq{eq:60}, yields the desired derivatives $\frac{d
  \phi_s(k)}{d\phi_l(0)}$.

\noindent {\it 3.} Lastly we derive an explicit form for
$\frac{d{F}}{du_m(k)}$, the desired performance gradient with 
respect to the physical control fields. From \Eq{eq:11}, we have
\begin{equation}
  \label{eq:36}
iH(k)=\sum_{l=1}^{8} a_lX_l+\sum_{l=1}^8 u_l(k)X_l,
\end{equation}
and $U_k=e^{-i H(k)\Delta t}$, where $k=0$, \dots, $N-1$. Define
\begin{eqnarray*}
\rho_k&=&U_{k-1} \cdots U_0 \rho_0 U_0^\dag \cdots U_{k-1}^\dag,\\
\Lambda_{k}&=& U_{k}^\dag \cdots U_{N-1}^\dag \rho_T U_{N-1}\cdots U_{k}.
\end{eqnarray*}
Then $\rho_N=\rho(T)$, $\Lambda_{N}=\rho_T$, and
\begin{eqnarray}
     \label{eq:37}
{F}&=&\tr \rho_T \rho(T)=\tr \Lambda_N \rho_N 
=\tr \Lambda_{N-1} \rho_{N-1}\nonumber\\
& =&\cdots =\tr \Lambda_1 \rho_1=\tr \Lambda_0 \rho_0.
\end{eqnarray}
It follows that
\begin{eqnarray}
  \label{eq:38}
&&\frac{d {F}}{d u_m(k)}= \frac{d\tr \Lambda_{k+1}\rho_{k+1}}{d u_m(k)}
= \frac{d\tr \Lambda_{k+1} U_k \rho_{k}U_k^\dag}{d u_m(k)}\nonumber\\
&=&\tr \Lambda_{k+1} \left(\frac{d U_k }{d u_m(k)}\rho_{k}U_k^\dag
+ U_k \rho_{k} \frac{dU_k^\dag}{d u_m(k)}\right).
\end{eqnarray}
From the following formula~\cite{Najfeld:95}
\begin{equation}
  \label{eq:39}
\left.\frac{d}{dv} e^{-i(H_a+vH_b)t}\right|_{v=0}
=-i \int_0^t e^{-i H_a\tau} H_b e^{i H_a\tau} d\tau\ e^{-i H_a t},
\end{equation}
we have
\begin{eqnarray}
  \label{eq:40}
  \frac{d U_k }{d u_m(k)}  
=- \int_0^{\Delta t} e^{-i H(k)\tau} X_m e^{i H(k)\tau} d\tau\ U_k.
\end{eqnarray}
Substituting \Eq{eq:40} into \eqref{eq:38}, we obtain
\begin{eqnarray*}
  \label{eq:41}
\frac{d {F}}{d u_m(k)} &=&\tr \Lambda_{k+1} \left(
- \int_0^{\Delta t} e^{-i H(k)\tau} X_m e^{i H(k)\tau} d\tau
\rho_{k+1}\right. \nonumber \\
&&\quad \left.+\rho_{k+1} \int_0^{\Delta t} e^{-i H(k)\tau} X_m e^{i H(k)\tau} d\tau
 \right)  \nonumber\\
&=& \tr [\Lambda_{k+1}, \rho_{k+1}]\int_0^{\Delta t} e^{-i H(k)\tau} X_m e^{i H(k)\tau}
d\tau.
\end{eqnarray*}
Since $H(k)$ is a Hermitian matrix, we can diagonalize it as
\begin{equation}
  \label{eq:69}
  H(k)=T(k) \Gamma(k) T^\dag(k),
\end{equation}
where $T(k)$ is a unitary matrix and $\Gamma(k)=\diag \{ \gamma_1,
\gamma_2, \gamma_3\}$.  Therefore,
\begin{eqnarray}
  \label{eq:21}
&&\int_0^{\Delta t} e^{-i H(k)\tau} X_m e^{i H(k)\tau}d\tau\nonumber\\
&=& \int_0^{\Delta t} T(k) e^{-i \Gamma(k)\tau} T^\dag(k) X_m T(k)
e^{i \Gamma(k)} T^\dag(k) d\tau\nonumber \\
&=&T(k) \int_0^{\Delta t} (T^\dag(k) X_m T(k))\odot {\Theta} d\tau\
T^\dag(k),
\end{eqnarray}
where $\odot$ denotes the Hadamard product, \ie, element-wise product, of
two matrices, and ${\Theta}_{ab}= e^{i(\gamma_b-\gamma_a)\tau}$. For
$\gamma_a\neq \gamma_b$, we define
\begin{eqnarray*}
  \label{eq:26}
\Phi_{ab}=  \int_0^{\Delta t} {\Theta}_{ab} d\tau
=\frac{e^{i(\gamma_b-\gamma_a)\Delta t}-1}
{i(\gamma_b-\gamma_a)};
\end{eqnarray*}
and for $\gamma_a=\gamma_b$, $\Phi_{ab}=\Delta t$. Therefore,
\begin{eqnarray*}
  \label{eq:24}
&&\int_0^{\Delta t} e^{-i H(k)\tau} X_m e^{i H(k)\tau}d\tau
\nonumber\\
&=& T(k) \big( (T^\dag(k) X_m T(k)) \odot \Phi\big) T^\dag(k),
\end{eqnarray*}
and
\begin{eqnarray}
  \label{eq:27}
\frac{d {F}}{d u_m(k)}&=&\tr ([\Lambda_{k+1}, \rho_{k+1}] 
 T(k)\nonumber\\
&& \quad\cdot\big((T^\dag(k) X_m T(k)) \odot \Phi\big) T^\dag(k).
\end{eqnarray}

We now have all the three factors in each term in the sum for the
desired performance gradient with respect to initial conditions,
$\frac{d{F}}{d\phi_l(0)}$, \Eq{eq:55}.

\section{Form of $\frac{d{F}}{d\phi_l(0)}$ for electron shuttling across triple quantum dot}
Now for the triple quantum dot system in Section~\ref{sec:lg_chen},
the gradient of the fidelity ${F}$ with respect to $\phi_l(0)$ is then
derived as
\begin{eqnarray*}
&&\frac{d{F}}{d\phi_l(0)}
=\sum_{k=0}^{N-1} \sum_{m\in\{7,8\}}  \sum_{s=1}^{8}
\frac{d{F}}{du_m(k)}\frac{d u_m(k)}{d\phi_s(k)}
\frac{d \phi_s(k)}{d\phi_l(0)}.\\
&=&\sum_{k=0}^{N-1} \left(
\frac14 \frac{d{F}}{du_7(k)}\frac{d \phi_7(k)}{d\phi_l(0)}
+\frac{1}{4\sqrt{3}}\frac{d{F}}{du_7(k)}\frac{d \phi_8(k)}{d\phi_l(0)}\right.\\
&&\left.+\frac{1}{4\sqrt{3}}\frac{d{F}}{du_8(k)}
\frac{d \phi_7(k)}{d\phi_l(0)}
+\frac{5}{12}\frac{d{F}}{du_8(k)}\frac{d \phi_8(k)}{d\phi_l(0)}\right).
\end{eqnarray*}
The Jacobian matrix $D{S}(\phi)$ (see \Eq{eq:59} for definition) 
is given in this case by 
\begin{equation*}
\ma{0&0&J_2&\substack{-\frac{\phi_7}2\\
-\frac{\sqrt{3} \phi_8}6}&0&0&-\frac{\phi_4}{2}&-\frac{\sqrt{3}\phi_4}6\\
0&0&-J_1&0&-\frac{\phi_8}{\sqrt{3}}&0&0&-\frac{\phi_5}{\sqrt{3}}\\
-J_2&J_1&0&0&0&\substack{\frac{\sqrt{3}\phi_8}2\\
+\frac{\phi_7}2}&\frac{\phi_6}{2}&\frac{\sqrt{3}\phi_6}2 \\
\substack{\frac{\sqrt{3}\phi_8}6\\
+\frac{\phi_7}2}&0&0&0&0&-J_2&\substack{\frac{\phi_1}2\\-2J_1}&\frac{\sqrt{3}\phi_1}6\\
0&\frac{\phi_8}{\sqrt{3}}&0&0&0&J_1&J_2& \substack{\frac{\phi_2}{\sqrt{3}}\\-\sqrt{3}J_2}\\
0&0&\substack{-\frac{\sqrt{3}\phi_8}2\\
-\frac{\phi_7}2}&J_2&-J_1&0&-\frac{\phi_3}2&-\frac{\sqrt{3}\phi_3}2 \\
0&0&0&2 J_1&-J_2&0&0&0\\
0&0&0&0&\sqrt{3}J_2&0&0&0}.
\end{equation*}

\section{Form of $\frac{d{F}}{d\phi_l(0)}$ for electron shuttling across ionized donor chain}
For the ionized donor chain of Section~\ref{sec:lg_greentree}, the
gradient of the fidelity ${F}$ with respect to $\phi_l(0)$ can then be
derived as
\begin{equation}
  \label{eq:66}
\frac{d{F}}{d\phi_l(0)}=
\sum_{k=0}^{N-1}\left(
\frac{d{F}}{du_1(k)}\frac{d \phi_1(k)}{d\phi_l(0)}
+\frac{d{F}}{du_2(k)}\frac{d \phi_2(k)}{d\phi_l(0)}
\right),
\end{equation}
and the Jacobian matrix $D{S}(\phi)$ is derived from \Eq{eq:53} as
 \begin{equation*}
\ma{ 0&\phi_3&\phi_2&\Delta&0&0&0&0\\
-\phi_3&0&-\phi_1&0&-\Delta&0&0&0\\
0&0&0&0&0&0&0&0\\
\substack{-\Delta\\-2\phi_7}&-\phi_6&0&0&0&-\phi_2&-2 \phi_1&0\\
\phi_6&\substack{\Delta+\phi_7\\-\sqrt{3} \phi_8}&0&0&0&\phi_1&\phi_2&-\sqrt{3}\phi_2\\
-\phi_5&\phi_4&0&\phi_2&-\phi_1&0&0&0\\
2 \phi_4&-\phi_5&0&2 \phi_1&-\phi_2&0&0&0\\
0&\sqrt{3} \phi_5&0&0&\sqrt{3}\phi_2 &0&0&0}.
\end{equation*}
For the particular Hamiltonian given in Eq.~\eqref{eq:34}, we can derive an
analytic solution for the decomposition in \Eq{eq:69}. The corresponding eigenvalues
of $H(k)$ are
\begin{equation*}
\gamma_1=-\frac{\Delta}3, \quad \gamma_2=\frac{\Delta+3g_1}6,\quad
 \gamma_3=\frac{\Delta-3g_1}6,
\end{equation*}
and the unitary matrix $T(k)$ is
\begin{equation*}
\ma{-\Omega_{23}/g_2 &
 \Omega_{12}/\sqrt{g_1(g_1+\Delta)/2}& \Omega_{12}/\sqrt{g_1(g_1-\Delta)/2}\\
0 & -\sqrt{(g_1+\Delta)/(2g_1)} & \sqrt{(g_1-\Delta)/(2g_1)} \\
\Omega_{12}/g_2 & \Omega_{23}/\sqrt{g_1 (g_1+\Delta)/2}
&\Omega_{23}/\sqrt{g_1(g_1-\Delta)/2)}},
\end{equation*}
where $g_1=\sqrt{\Delta^2+4\Omega_{23}^2+4\Omega_{12}^2}$ and
$g_2=\sqrt{\Omega_{23}^2+\Omega_{12}^2}$.

\bibliographystyle{apsrev}

\end{document}